\documentclass[aps,pre,reprint,superscriptaddress,amsmath,amssymb,showpacs,floatfix,longbibliography]{revtex4-1}

\usepackage{graphicx}
\usepackage{dcolumn}
\usepackage{bm}
\usepackage{color}
\usepackage[utf8]{inputenc}
\usepackage{physics}
\usepackage{mathtools}
\usepackage{soul}

\begin{document}
\title{Quantum mechanical bound for efficiency of quantum Otto heat engine}
\author{Jong-Min Park}
\affiliation{School of Physics, Korea Institute for Advanced Study, Seoul
02455, Korea}
\affiliation{Department of Physics, University of Seoul, Seoul 02504, Korea}
\author{Sangyun Lee}
\affiliation{Department of Physics, Korea Advanced Institute of Science and Technology, Daejeon 34141, Korea}
\author{Hyun-Myung Chun}
\affiliation{II. Institut f{\"u}r Theoretische Physik, Universit{\"a}t Stuttgart, 70550 Stuttgart, Germany}
\author{Jae Dong Noh}
\affiliation{Department of Physics, University of Seoul, Seoul 02504, Korea}

\date{\today}

\begin{abstract}
{
The second law of thermodynamics constrains that the efficiency of heat engines,
classical or quantum, cannot be greater than the universal Carnot efficiency.
We discover another bound for the efficiency of a quantum Otto heat engine 
consisting of a harmonic oscillator.
Dynamics of the engine is governed by the Lindblad
equation for the density matrix, which is mapped to the Fokker-Planck
equation for the quasi-probability distribution. Applying stochastic
thermodynamics to the Fokker-Planck equation system, we obtain the
$\hbar$-dependent quantum 
mechanical bound for the efficiency. It turns out that the bound 
is tighter than the Carnot efficiency.
The engine achieves the bound in the low 
temperature limit where quantum effects dominate. Our work demonstrates that
quantum nature could suppress the performance of heat engines in terms of efficiency bound, work and power output.
}
\end{abstract}

\maketitle

\section{Introduction}\label{sec1}
A heat engine is a device harvesting work making use of a heat flow 
between multiple thermal reservoirs.
One of the main concerns for the heat engine is efficiency.
When the heat engine is in contact with two thermal reservoirs at
temperatures $T_1$
and $T_2(< T_1)$, the second law of thermodynamics constrains that the
efficiency cannot be greater than the Carnot efficiency 
$\eta_C = 1 - T_2/T_1$~\cite{Kittel:1980tpb}.
The upper bound is universal and independent of specific properties 
of heat engines.

We address the question of whether the Carnot efficiency is the unique 
fundamental bound for a quantum heat engine, a heat engine whose working
substance is governed by quantum mechanics~\cite{alicki1979quantum,kosloff1984quantum,uzdin2015equivalence}.
Suppose that the temperature is so low that the thermal energy is 
comparable to or even less than the relevant energy scale.
Then, quantum mechanical effects may show up and be reflected 
in the efficiency and its bound.
Various quantum heat engine models have been studied to find the traces of
quantum effects.
On the one hand, some quantum heat engines behave similarly to classical
engines as far as they are in contact with thermal
reservoirs~\cite{uzdin2015equivalence, kosloff2017quantum}: 
the efficiency is bounded by the Carnot efficiency from 
above~\cite{alicki1979quantum,lin2003performance,bender2000quantum,kieu2006quantum,kieu2004second}
and the efficiency at the maximum power is given by the 
Curzon-Ahlborn
efficiency~\cite{kosloff2017quantum,kosloff1984quantum,lin2003performance,Abah:2012Single-ion}.
On the other hand, coherence and entanglement effects
have been observed in quantum engines in contact with nonequilibrium 
reservoirs~\cite{rossnagel2014nanoscale,scully2003extracting,abah2014efficiency,Wolfgang:2016operation}
or with non-commutative operations~\cite{Diaz:2014Quantum-information, Kosloff:2002Discrete}. 

In this paper, we investigate the quantum mechanical bound for 
the efficiency of the quantum Otto heat engine which uses a simple harmonic
oscillator as a working substance~\cite{Abah:2012Single-ion,abah2014efficiency,Jiawen:2013Boosting,Campo:2014more,Wolfgang:2016operation}.
The quantum Otto heat engine has gathered more attention as it became
realizable experimentally~\cite{Abah:2012Single-ion}.
The quantum mechanical state of the engine is described by the density
matrix. We find that the quasi-probability distribution representation of
the density matrix is useful~\cite{santos2017wigner}.
The equation of motion for the density matrix can be mapped to a classical
Fokker-Planck equation for the quasi-probability 
distribution~\cite{gardiner2004quantum}. 
By applying stochastic thermodynamics to
the effective Fokker-Planck equation~\cite{seifert:2005Entropy}, we obtain the
$\hbar$-dependent upper bound for the engine efficiency. 
Interestingly, the bound is tighter than the Carnot efficiency. Our work
elucidates that the quantum mechanical effects could suppress the performance 
of heat engines in terms of efficiency bound, work and power output.

This paper is organized as follows. We introduce the quantum Otto heat
engine model in Sec.~\ref{sec:model}. 
The engine cycle consists of the adiabatic and isochoric processes. Dynamics
of the density operator during the processes is described. 
In Sec.~\ref{sec:qpd}, we
introduce the quasi-probability distribution and derive the equation of
motion for it. The quasi-probability distribution satisfies the
Fokker-Planck equation, to which one can apply the classical thermodynamics.
In Sec.~\ref{sec:FP}, we derive the quantum mechanical bound for the engine
efficiency by analyzing the Fokker-Planck equation system. 
Quantum mechanical effects on the heat engine are discussed in
Sec.~\ref{quantum_bound}. We conclude the paper with summary and discussions
in Sec.~\ref{conclusion}.

\section{Quantum Otto heat engine}\label{sec:model}

We consider a quantum Otto heat engine model with a simple harmonic
oscillator as a working substance. 
The system Hamiltonian $\hat{H}(t)$ is given by
\begin{equation}\label{eq:Hamiltonian}
    \hat{H}(t) = \hbar \omega(t) \hat{a}^\dagger\hat{a},
\end{equation}
where $\hat{a}^\dagger$ and $\hat{a}$ are the creation and the annihilation operators
satisfying $\comm*{\hat{a}}{\hat{a}^\dagger} = 1$. The frequency parameter
$\omega(t)$ is varied cyclically in time in a prescribed manner.
The quantum mechanical state of the system is described by the density
operator $\hat{\rho}(t)$.
The same Hamiltonian was studied to find the optimal 
condition for the quantum heat engine operating in the Carnot 
cycle~\cite{lin2003performance, Wang:2007Performance,Liu2009:Ecological}.

\begin{figure}
\includegraphics*[width=\columnwidth]{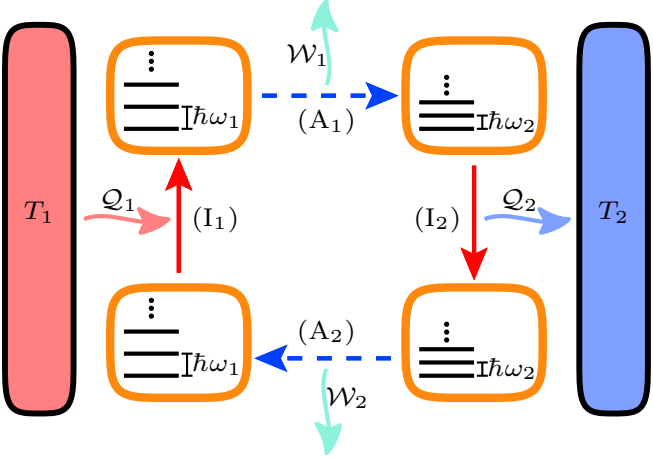}
\caption{
Illustration of the engine cycle.
The energy levels of the harmonic oscillator are depicted inside the orange
boxes.
The solid~(red) and the dashed~(blue) arrows represent the isochoric processes and
the adiabatic processes, respectively. The system is in contact with the
thermal reservoirs during the isochoric processes.
The wavy arrows indicates the direction of work and heat flows.
} 
\label{fig_engine}
\end{figure}

Our engine system operates in the Otto cycle consisting of adiabatic
and isochoric processes as illustrated in Fig.\ref{fig_engine}.
During the adiabatic process, the system is isolated from the heat reservoir
and the frequency parameter $\omega(t)$ varies in time $t$ between
$\omega_1$ and $\omega_2 (<\omega_1)$.
We denote the adiabatic processes starting with $\omega_1$ and $\omega_2$
by ${\rm A}_1$ and $\rm{A}_2$, respectively.
The density matrix is governed by the von Neumann 
equation~\cite{gardiner2004quantum}
\begin{equation}\label{vNeq}
    \pdv{t} \hat{\rho} (t) = -\frac{i}{\hbar} \comm{\hat{H} (t)}{\hat{\rho}
(t)} .
\end{equation}

During the isochoric process, 
the system is connected to the thermal reservoir of temperature $T_i$ while
the frequency parameter is kept constant at $\omega_i~(i=1,2)$.
These isochoric processes are denoted as ${\rm I}_1$ and ${\rm
I}_2$, respectively.
We adopt the Lindblad master equation to describe the dynamics
during the isochoric process. 
The Lindblad master equation~\cite{breuer2002theory} during the process
${\rm I}_i$ is given by
\begin{equation}\label{Lindblad_Eq}
    \pdv{t} \hat{\rho} (t) = -\frac{i}{\hbar} \comm{\hat{H}_i}{\hat{\rho}
(t)} + \mathcal{D}_i \qty(\hat{\rho}(t)) 
\end{equation}
with the dissipator $\mathcal{D}_{i}$ defined by
\begin{align}\label{dissipator}
    \mathcal{D}_i \qty(\hat{\rho}) = &\gamma \qty(\bar{n}_i + 1) \qty[\hat{a} \hat{\rho} \hat{a}^\dagger
    - \frac{1}{2} \acomm{\hat{a}^\dagger \hat{a}}{\hat{\rho}} ] \nonumber\\
    &+ \gamma \bar{n}_i \qty[\hat{a}^\dagger \hat{\rho} \hat{a}
    - \frac{1}{2} \acomm{\hat{a} \hat{a}^\dagger}{\hat{\rho}} ].
\end{align}
Here, $\gamma$ is a damping rate and
\begin{equation}
\bar{n}_i = \left(e^{\beta_i \hbar \omega_i}-1\right)^{-1} 
\end{equation}
is the Planck distribution at inverse 
temperature $\beta_i = 1/(k_B T_{i})$. The Boltzmann constant $k_B$ will be
set to unity.
The Lindblad equation has the thermal equilibrium state 
\begin{equation}\label{rho_th}
    \hat{\rho}_{\rm th} =  \qty(1-e^{-\beta_i\hbar\omega_i} )
e^{-\beta_i \hat{H}_i} 
\end{equation}
as its steady state solution.

It takes $t_m$ for each process $m={\rm A}_1,{\rm A}_2,{\rm I}_1$, and
${\rm I}_2$ so that the total engine cycle time is 
$\tau = t_{{\rm A}_1} + t_{{\rm A}_2} + t_{{\rm I}_1} + t_{{\rm I}_2}$.
Repeating the cycles, the system will reach the cyclic steady
state. We find that the density matrix in the cyclic steady state is 
of the form
\begin{equation}\label{cyclic_steady}
\hat{\rho}(t) = \qty(1-e^{-c(t)}) e^{-c(t) \hat{a}^\dagger\hat{a}}
\end{equation}
with a periodic function $c(t) = c(t+\tau)$. 
In this state, the expectation value 
of the number operator $\hat{N} = \hat{a}^\dagger \hat{a}$ 
is given by $N(t) \equiv \tr{ \hat{N} \hat{\rho}(t)}
= \qty(e^{c(t)}-1)^{-1}$.
Thus, the cyclic steady state is fully characterized by $N(t) = N(t+\tau)$.

\begin{figure}
\includegraphics*[width=\columnwidth]{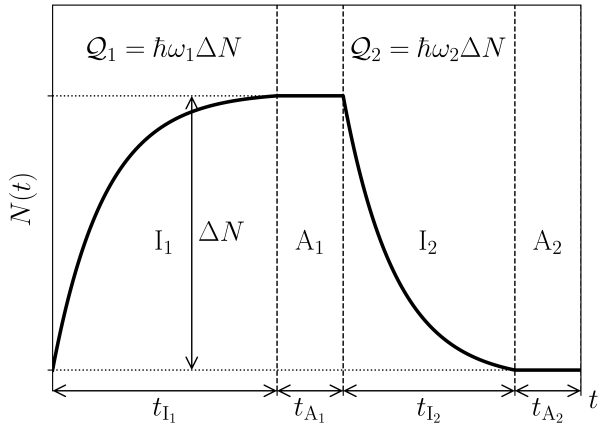}
\caption{
Schematic plot of $N(t)$ over an engine cycle.
} 
\label{fig_DeltaN}
\end{figure}

The von Neumann equation \eqref{vNeq} yields that $N(t) = N_i$ is
a time-independent constant during the adiabatic process ${\rm A}_i$. 
On the other
hand, during the isochoric process ${\rm I}_i$, the Lindblad 
equation~\eqref{Lindblad_Eq} yields that
\begin{equation}\label{dNdt}
\frac{d}{dt}N(t) = - \gamma \qty( N(t) - \bar{n}_i ) .
\end{equation}
The solution $N(t) = \bar{n}_i+ \qty( N(t_0) -\bar{n}_i) 
e^{-\gamma(t-t_0)}$ 
provides a self-consistent equation for $N_1$ and $N_2$, which leads to 
\begin{equation}\label{DeltaN}
\Delta N = N_1 - N_2 =  
    \qty( \bar{n}_1 - \bar{n}_2 )
    \frac{\qty( 1 - e^{-\gamma t_{{\rm I}_1}} )
    \qty( 1 - e^{-\gamma t_{{\rm I}_2}} )}{
    1 - e^{-\gamma ( t_{{\rm I}_1} + t_{{\rm I}_2} )}}.
\end{equation}
Time-dependence of $N(t)$ over the engine cycle is plotted in
Fig.~\ref{fig_DeltaN}.

The expectation value of the internal energy 
$\mathcal{E}(t) = \tr \{ \hat{H} (t) \hat{\rho} (t)\}$
varies in time at the rate
\begin{equation}\label{energy_rate}
    \dot{\mathcal{E}} = \tr{ \qty( \pdv{t} \hat{H} (t) ) \hat{\rho} (t) }
    + \tr{ \hat{H} (t) \qty( \pdv{t} \hat{\rho} (t) ) },
\end{equation}
where the former~(latter) is designated as the rate for 
work~(heat)~\cite{alicki1979quantum, kosloff1984quantum}.
During the isochoric processes, the Hamiltonian is time-independent and the
system absorbs or dissipates the heat performing no work. During the
adiabatic processes, the system performs the work without heat exchange.
We will use the sign convention for the work and heat as specified in
Fig.~\ref{fig_engine}.

Since the internal energy is given by $\mathcal{E}(t) = \hbar\omega(t) N(t)$,
the heat and the net work $\mathcal{W} = \mathcal{W}_1+\mathcal{W}_2$ 
for the single engine cycle are written in terms of
$\Delta N$ as 
\begin{equation}\label{Q1_Q2_W}
\begin{split}
{\cal Q}_1 &=  \hbar\omega_1 \Delta N,  \\
{\cal Q}_2 &=  \hbar\omega_2 \Delta N,  \\
{\cal W}   &=  \hbar (\omega_1-\omega_2)\Delta N  .
\end{split}
\end{equation}
Note that $\mathcal{W} = \mathcal{Q}_1 - \mathcal{Q}_2$,
which corresponds to the first law of thermodynamics.
The system acts as a heat engine when $\mathcal{Q}_1 \geq 0$ and
$\mathcal{W} \geq 0$ or $(\Delta N)\geq0$ and $w_1\geq w_2$.
Then, the efficiency is given by
\begin{equation}\label{eta}
    \eta = \frac{\mathcal{W}}{\mathcal{Q}_1} = 
           1 - \frac{\mathcal Q_2}{\mathcal Q_1} = 
           1 - \frac{\omega_2}{\omega_1}.
\end{equation}
The condition $\Delta N\geq 0$ requires that $\omega_1/T_1 \leq
\omega_2/T_2$~(see \eqref{DeltaN}). Consequently, the engine efficiency
cannot be larger than the Carnot efficiency
\begin{equation}\label{Carnot_eff}
\eta_C = 1 - \frac{T_2}{T_1} .
\end{equation}

The Carnot efficiency is also derived from the thermodynamic principle.
Consider the von Neumann entropy
\begin{equation}
S_{\rm vN}(t) = - \tr{ \hat{\rho}(t) \ln\hat{\rho}(t)} .
\end{equation}
Over the isochoric processes governed by the Lindblad
equation~\eqref{Lindblad_Eq}, the system
should satisfy the second law of thermodynamics~\cite{Spohn:1978Irreversible}
\begin{equation}\label{2nd_law_Lindblad}
\begin{split}
  \qty(\Delta S_{\rm vN})_{{\rm I}_1} - \frac{\mathcal{Q}_1}{T_1} & \geq 0 , \\
  \qty(\Delta S_{\rm vN})_{{\rm I}_2} + \frac{\mathcal{Q}_2}{T_2} & \geq 0 .
\end{split}
\end{equation}
On the other hand, the entropy is invariant
$\qty(\Delta S_{\rm vN})_{{\rm A}_1}= \qty(\Delta S_{\rm vN})_{{\rm
A}_2}=0$ during the adiabatic processes~\cite{gardiner2004quantum}.
The von Neumann entropy changes over the entire engine cycle add up 
to be zero. Consequently,
\eqref{2nd_law_Lindblad} leads to the inequality
\begin{equation}\label{second_law}
    -\frac{\mathcal{Q}_1}{T_1} + \frac{\mathcal{Q}_2}{T_2} \geq 0 
\end{equation}
and the Carnot bound.

We remark that the Carnot efficiency is the universal bound irrespective of
system-dependent details. 
The same thermodynamic bound was also found in the previous studies 
of the quantum Otto
heat engine~\cite{kosloff2017quantum,kieu2006quantum,kieu2004second}.
It may suggest that the quantum mechanical nature does not impose
an additional constraint on the efficiency. 
In the following section, however, we will discover another bound for the
efficiency that is tighter than the Carnot efficiency.

\section{Quasi-Probability distribution}\label{sec:qpd}

The quasi-probability distribution allows a semi-classical description of
a quantum mechanical system~\cite{gardiner2004quantum}.
Recently, the quasi-probability distribution proved to be useful for the
study of thermodynamics of open quantum
systems~\cite{santos2017wigner,santos2018spin,santos2018irreversibility}. We
investigate the quantum Otto heat engine
using the quasi-probability distribution.

The quasi-probability distributions can be defined by
the Fourier transform of a joint moment generating function
of $\hat{a}^\dagger$ and $\hat{a}$~\cite{gardiner2004quantum}.
Unlike the probability distributions for classical observables, the 
quasi-probability distributions do not have a unique representation 
due to the nonvanishing commutator of the operators.
Most commonly studied are the P-representation $P(\alpha,\alpha^*)$, 
the Husimi Q-distribution $Q(\alpha,\alpha^*)$, and the Wigner function 
$W(\alpha,\alpha^*)$~\cite{gardiner2004quantum}.
In this paper, we present the results mainly from the Husimi Q-distribution
or the Q-function. The results from the other distributions will be
mentioned briefly.

Let $|\alpha\rangle$ and $\langle \alpha|$ be the coherent states
satisfying $\hat{a}|\alpha\rangle =
\alpha|\alpha\rangle$ and $\langle \alpha| \hat{a}^\dagger = \alpha^* \langle
\alpha|$ with a complex number $\alpha$ and its complex conjugate
$\alpha^*$.
The Q-function is defined as~\cite{husimi1940some,gardiner2004quantum}
\begin{equation}\label{eq:Q-function}
    Q (\alpha, \alpha^*, t) = \frac{1}{\pi} \ev{\hat{\rho} (t)}{\alpha}.
\end{equation}
It allows one to evaluate moments of $\hat{a}$ and $\hat{a}^\dagger$ 
in the antinormal order conveniently as~\cite{gardiner2004quantum}
\begin{equation*}
\langle \hat{a}^m
\hat{a}^{\dagger n}\rangle = \tr{ \hat{\rho}(t) \hat{a}^m
\hat{a}^{\dagger n}} = \int d^2\alpha~ \alpha^m\alpha^{* n}
Q(\alpha,\alpha^*,t) .
\end{equation*}

All the expressions involving the density operator can be rewritten 
in terms of the Q-function. Mathematical tools for that purpose are found in
the literature. Thus, we present the following relations without derivation. 
We refer readers to Ref.~\cite{gardiner2004quantum} for details. 
First of all, the von Neumann equation \eqref{vNeq} becomes
\begin{equation}\label{ev_Q_adi}
\frac{\partial}{\partial t}Q(\alpha,\alpha^*,t) =
i \omega(t) \frac{\partial}{\partial \alpha} \alpha Q - i 
\omega(t) \frac{\partial}{\partial \alpha^*} \alpha^* Q .
\end{equation}
It has the solution
\begin{equation}\label{Q_adiabatic}
Q(\alpha,\alpha^*,t) = Q_0\qty( \alpha e^{i\Omega(t)},\alpha^*
e^{-i\Omega(t)}) ,
\end{equation}
where $Q_0(\alpha,\alpha^*)$ is the initial distribution at time $t_0$ and 
$\Omega(t) \equiv \int_{t_0}^t \omega(t') dt'$. 
It can be easily derived
from the definition $Q(t) = \frac{1}{\pi}\langle \alpha |
\hat{\rho}(t) | \alpha\rangle$. 
Note that $\hat{\rho}(t) = \hat{U}(t)
\hat{\rho}(t_0) \hat{U}^\dagger(t)$ with the unitary time evolution 
operator $\hat{U}(t) = e^{-i \Omega(t) \hat{N}}$. 
The identity $e^{-i\Omega \hat{N}} \hat{a} e^{i\Omega\hat{N}} =
e^{i\Omega} \hat{a}$ leads to 
$\hat{U}^\dagger(t) |\alpha\rangle \propto |\alpha e^{i \Omega(t)}\rangle$.
Thus, during the adiabatic process, 
the Q-function rotates in the complex $\alpha$ plane by the angle
$\Omega(t)$ maintaining its shape.

The Lindblad equation ~\eqref{Lindblad_Eq} for the isochoric process 
is rewritten as~\cite{gardiner2004quantum}
\begin{equation}\label{ev_Q_iso}
    \pdv{t} Q(\alpha,\alpha^*,t) = - 
\qty( \pdv{\alpha} J + \pdv{\alpha^*} J^* ) ,
\end{equation}
where $J$, which will be called the probability current, is given by
\begin{equation}\label{eq:current}
    J
    = - \qty( i \omega + \frac{\gamma}{2} ) \alpha Q
    - D \pdv{\alpha^*} Q
\end{equation}
and $D = \gamma ( \bar{n} + 1 )/2$, called the diffusion constant.
Note that $\omega = \omega_i$ and $\bar{n} = \bar{n}_i$ for the adiabatic
process ${\rm I}_i$~$(i=1,2)$.
We remark that \eqref{ev_Q_iso} also covers the adiabatic process
when one sets $\gamma = 0$ and replaces $\omega$ with 
the time-dependent $\omega(t)$.
Thus, we can use the equation of motion \eqref{ev_Q_iso} to describe both 
the adiabatic and isochoric processes.
The other quasi distributions have the same equations of motion with
their own diffusion constants.
The P-representation has $D = \gamma \bar{n}/2$ and the Wigner function has 
$D = \gamma(\bar{n} + 1/2)/2$.

The thermal equilibrium state \eqref{rho_th} is rewritten as 
\begin{equation}\label{Qss}
Q_{\rm th}(\alpha,\alpha^*) = \frac{1}{\pi (\bar{n}+1) }
  e^{-\alpha\alpha^* /(\bar{n}+1)} ,
\end{equation}
while the cyclic steady state solution~\eqref{cyclic_steady} becomes 
\begin{equation}
Q(\alpha,\alpha^*,t) = \frac{1}{\pi(N(t)+1)} e^{-\alpha\alpha^* /
(N(t)+1)} .
\end{equation}
They are obtained by using the identity $e^{-\lambda
\hat{a}^\dagger \hat{a}} =
\colon\!{e^{-(1-e^{-\lambda})\hat{a}^\dagger\hat{a}}}\colon$
where $\colon\! \hat{O}\colon$ represents the normal 
ordered form of an operator $\hat{O}$~\cite{Blasiak:2007Combinatorics}.

The expectation value of the number operator 
is also rewritten in terms of the Q-function:
\begin{equation}
\begin{split}
N(t) &= \tr{ \hat{a}^\dagger\hat{a}\hat{\rho} } 
      = \tr{ (\hat{a}\hat{a}^\dagger -1)\hat{\rho} } \\
     &= \int d^2\alpha~ \left[(\alpha\alpha^*-1) Q\right]  .
\end{split}
\end{equation}
The internal energy and the heat absorption rate are written similarly as
\begin{equation}\label{energy_Q}
{\mathcal E} = \hbar\omega \int d^2\alpha~ \left[(\alpha\alpha^*-1)
Q\right] \\ 
\end{equation}
and 
\begin{equation}\label{heatQ}
\begin{split}
\dot{\mathcal Q} &= \hbar\omega \int d^2\alpha~ \left[(\alpha\alpha^*-1) 
  \frac{\partial Q}{\partial t}\right]  \\ 
  &= \gamma \hbar\omega \int d^2\alpha~ \qty[(\bar{n}+1) -\alpha\alpha^*]~ Q .
\end{split}
\end{equation}
The last equality is obtained by using \eqref{ev_Q_iso}.

\section{Fokker-Planck equation and thermodynamics}\label{sec:FP}

The quasi-probability distribution $Q(\alpha,\alpha^*)$ 
is a real-valued nonnegative and normalized function. 
Furthermore, for the harmonic oscillator system, 
the second-order partial differential equation for $Q$ 
as shown in~\eqref{ev_Q_iso} has
the same structure as the Fokker-Planck equation for a classical
Markov system. We exploit the correspondence to 
map the quantum Otto heat engine to a classical thermodynamic system.

Consider first the isochoric process.
We introduce a {\em position-like} variable $x = (\alpha+\alpha^*)/2$ and a
{\it momentum-like} variable $p=(\alpha-\alpha^*)/(2i)$. 
Then, the Lindblad equation \eqref{ev_Q_iso} is rewritten as
\begin{equation}
\frac{\partial}{\partial t}Q(x,p,t) = - 
\frac{1}{\tilde{\gamma}} \sum_{k=x,p} \qty( \partial_k A_k 
- \tilde{T} \partial_k^2) Q
\end{equation}
where $\partial_k$ denotes the partial differentiation with respect to 
$k=(x, p)$, the drift force $A_k(x,p)$ is given by
\begin{equation}\label{drift_force}
\begin{pmatrix} A_x \\  A_p \end{pmatrix}
= \begin{pmatrix} - 2\hbar \omega  & \omega\tilde{\gamma}  \\
                  -\omega\tilde{\gamma} & - 2\hbar\omega   \end{pmatrix} 
\begin{pmatrix} x \\ p \end{pmatrix} ,
\end{equation}
and the parameters are given by
\begin{equation}\label{effective_parameter}
\begin{aligned}
\tilde{\gamma} &= \frac{4\hbar\omega}{\gamma} \\
\tilde{T} & = (\bar{n}+1)\hbar\omega  =
\frac{\hbar\omega}{1-e^{-\beta\hbar\omega}} .
\end{aligned}
\end{equation}
This is equivalent to the Fokker-Planck equation for a Brownian particle 
in the two-dimensional phase space $(x,p)$ under the drift force $A_k(x,p)$. 
The particle is  immersed in the
thermal reservoir characterized by the effective damping coefficient
$\tilde{\gamma}$ and the effective temperature $\tilde{T}$. The drift force
$A_k$ are linear in $x$ and $p$. Such a linear system is called  
the Ornstein-Uhlenbeck process, whose
properties are well documented in the 
literature~\cite{Gardiner:2010stochastic,Risken:1996FokkerPlanck,VanKampen:2011Stochastic}.

We are at liberty to assume that the momentum-like variable $p$ is odd under
the time reversal while $x$ is even. Following 
Ref.~\cite{Gardiner:2010stochastic}, one can show that the dynamics
satisfies the detailed balance. 
Thus, the Fokker-Planck equation describes an equilibrium system. 
The distribution function in \eqref{Qss} corresponds to the
equilibrium Boltzmann distribution $Q_{\rm th}(x,p) =
\frac{1}{Z} e^{-\tilde{\beta} V(x,p)}$, where
\begin{equation}\label{energy_V}
V(x,p) = \hbar\omega (x^2+p^2-1)
\end{equation}
is the energy function,
$\tilde{\beta} = 1/ \tilde{T}$ is the effective inverse 
temperature, and $Z$ is the partition function.
Due to the choice $\tilde{T} = (\bar{n}+1)\hbar\omega$, we have the
equivalence
\begin{equation}\label{e_equiv}
\mathcal{E}(t) = \int dx dp V(x,p)Q(x,p,t)
\end{equation}
between the energy expectation value of the quantum system
and the ensemble average of the energy function $V(x,p)$ 
of the effective classical system.

The same Fokker-Plank equation 
with $\gamma = 0$ and $\omega=\omega(t)$
covers the adiabatic process. 
The system is detached from the heat reservoir and
driven out of equilibrium with the time-dependent $\omega(t)$. 

We are now ready to apply classical thermodynamics to the Fokker-Planck
system. The second law of thermodynamics for the Fokker-Planck system states
that~\cite{seifert:2005Entropy}
\begin{equation}\label{2nd_law_FP}
\Delta S_{\rm tot} = 
\Delta S_Q + \frac{-\Delta \mathcal{Q}}{\tilde{T}} \geq 0 ,
\end{equation}
where $\Delta S_{Q}$ is the change in the Shannon entropy
\begin{equation}
\begin{aligned}
S_Q(t) &= -\int d^2\alpha~ Q(\alpha,\alpha^*,t) \ln Q(\alpha,\alpha^*,t) \\
       &= -\int{dx}{dp}~ Q(x,p,t) \ln Q(x,p,t)
\end{aligned}
\end{equation}
of the system and $-(\Delta \mathcal{Q})/\tilde{T}$ is the Clausius entropy
change of the heat reservoir of temperature $\tilde{T}$ 
losing the heat $(\Delta \mathcal{Q})$.
The Shannon entropy for the quasi-probability 
distribution is called the
Wehrl entropy~\cite{wehrl1979relation,wehrl1978general}. 
Due to the equivalence~\eqref{e_equiv}, the heat dissipations in the quantum
and the classical systems are the same. 
On the other hand, the Wehrl entropy, in general, 
is different from the von Neumann entropy 
$S_{\rm vN} = -\tr{\hat{\rho}\ln\hat{\rho}} = -\frac{1}{\pi} \int d^2\alpha 
\langle \alpha| \hat{\rho}\ln\hat{\rho} |\alpha\rangle$ which involves
$\langle \alpha | \hat{\rho} | \alpha'\rangle $ with $\alpha'\neq \alpha$.
Thus, the inequality in \eqref{2nd_law_FP} for the effective system
may provide an additional information that is unavailable from 
the second law \eqref{second_law} for the quantum system.

Applying the second law of thermodynamics to the effective system,
one obtains the following relations:
\begin{equation}
\begin{split}
(\Delta S_{\rm tot})_{{\rm I}_1} & = (\Delta S_Q)_{{\rm I}_1} -
\frac{\mathcal{Q}_1}{\tilde{T}_1} \geq 0  , \\
(\Delta S_{\rm tot})_{{\rm I}_2} & = (\Delta S_Q)_{{\rm I}_2} + 
\frac{\mathcal{Q}_2}{\tilde{T}_2} \geq 0  , \\
(\Delta S_{\rm tot})_{{\rm A}_1} &=(\Delta S_{\rm tot})_{{\rm A}_2} =0 .
\end{split}
\end{equation}
During the adiabatic processes, the total entropy does not change since 
the shape of $Q(x,p)$ is invariant~(see \eqref{Q_adiabatic}) 
and there are no heat dissipations. 
Since the Wehrl entropy is a state function, the sum of the Wehrl entropy
changes over the complete engine cycle adds up to zero. 
Therefore, we obtain
\begin{equation}\label{Q_inequality}
-\frac{{\mathcal Q}_1}{\tilde{T}_1} + \frac{{\mathcal Q}_2}{\tilde{T}_2}
\geq 0 .
\end{equation}
This inequality yields that the engine efficiency is bounded above by the
bound 
\begin{equation}\label{eta_Q}
\eta_Q = 1 - \frac{\tilde{T}_2}{\tilde{T}_1} = 1 - \frac{\omega_2
(1-e^{-\beta_1\hbar\omega_1})}{\omega_1(1-e^{-\beta_2\hbar\omega_2})} .
\end{equation}
This bound is different from the Carnot efficiency $\eta_C = 1-T_2/T_1$.
We will call this bound the quantum mechanical bound as it depends
explicitly on the Planck constant.

\section{Quantum mechanical effect}\label{quantum_bound}

We discuss the implication of the quantum mechanical bound $\eta_Q$.
In order to quantify the
quantum mechanical effect, we introduce a dimensionless parameter
\begin{equation}
q = \frac{\hbar\omega_1}{k_B T_1}.
\end{equation} We also introduce positive dimensionless parameters 
$r_T =  T_2/T_1$, $r_\omega = \omega_2/\omega_1$, and $r = r_\omega/r_T = 
\beta_2\omega_2/(\beta_1 \omega_1)$.
We only consider the region $r_T \leq 1$, $r_\omega \leq 1$, and $r \geq 1$ 
where the system acts as a heat engine.
The quantum
mechanical bound is then written as
\begin{equation}
\eta_Q = 1 - r_\omega \qty( \frac{1-e^{-q}}{1-e^{-r q}} ) .
\end{equation}

\begin{figure}
\includegraphics*[width=\columnwidth]{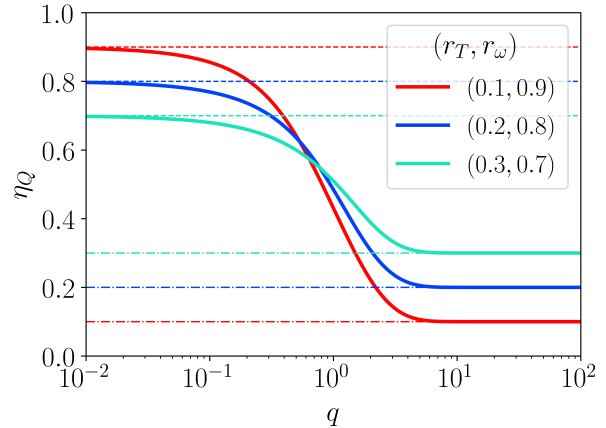}
\caption{
The quantum mechanical bounds $\eta_Q$ are plotted as
a function of $q$ for dimensionless parameters 
$(r_T,r_\omega)=(0.1,0.9)$ (red),
$(0.2,0.8)$ (blue), and $(0.3,0.7)$ (cyan).
The dashed line and the dashed-dotted line indicate
the Carnot efficiency $\eta_C$ and the efficiency $\eta$, respectively.
} 
\label{fig_eta_Q}
\end{figure}

The bound $\eta_Q$ is a decreasing function of $q$ and 
equal to the Carnot efficiency at $q=0$. Thus, we conclude that
\begin{equation}
\eta \leq \eta_Q \leq \eta_C .
\end{equation}
The quantum mechanical bound is tighter than the Carnot efficiency.
It reduces to the Carnot efficiency in the limiting case
$q\to 0$~(classical limit) or $r\to 1$~(reversible limit).
The $q$-dependence of $\eta_Q$ is drawn in Fig.~\ref{fig_eta_Q} for a couple
of values of $(r_T,r_\omega)$.

The Carnot efficiency is realized~($\eta = \eta_C$) 
in the reversible limit $r\to 1$. On the other hand, the quantum mechanical 
bound is realized~($\eta=\eta_Q$) in the $q\to \infty$ limit.
Thus, the quantum mechanical bound $\eta_Q$ is more useful than the 
Carnot efficiency $\eta_C$ as a fundamental bound for the efficiency.

\begin{figure}
\includegraphics*[width=\columnwidth]{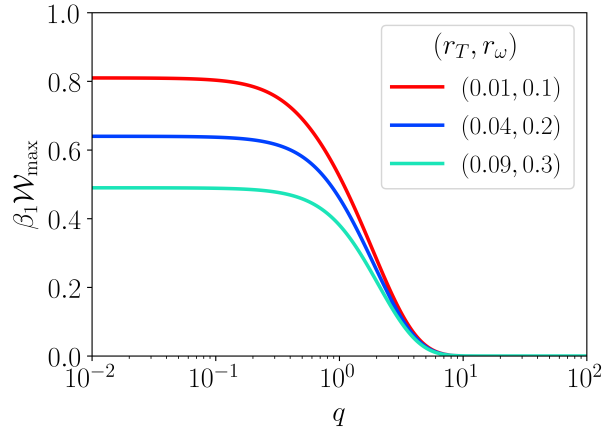}
\caption{
The maximum value of the work ${\mathcal W}_{\rm max}$
is plotted as a function of $q$
for dimensionless parameters $(r_T,r_\omega)=(0.01,0.1)$ (red),
$(0.04,0.2)$ (blue), and $(0.09,0.3)$ (cyan).
}
\label{fig_work_max}
\end{figure}

It is also interesting to study a quantum mechanical effect on the power of the
engine. From \eqref{DeltaN} and \eqref{Q1_Q2_W}, 
the extracted work per engine cycle is given by
\begin{equation}\label{WperCycle}
    {\mathcal W} =
    {\mathcal W}_{\rm max} \frac{\qty( 1 - e^{-\gamma t_{{\rm I}_1}} )
    \qty( 1 - e^{-\gamma t_{{\rm I}_2}} )}{
    1 - e^{-\gamma ( t_{{\rm I}_1} + t_{{\rm I}_2} )}} , 
\end{equation}
where 
\begin{equation}\label{Wmax}
\begin{split}
{\mathcal W}_{\rm max} &= 
\hbar (\omega_1 - \omega_2) ( \bar{n}_1 - \bar{n}_2 ) \\
&=
(1-r_\omega)k_B T_1  \qty(\frac{q}{e^{q}-1}-\frac{q}{e^{rq}-1}) .
\end{split}
\end{equation}
As a function of the cycle times, it takes the maximum value 
$\mathcal{W}_{\rm max}$ when $t_{{\rm I}_1} =  t_{{\rm I}_2} \to \infty$.
After a little algebra, one can show that ${\mathcal W}_{\rm max}$ is a
decreasing function of $q$~(see Fig.~\ref{fig_work_max}). 
It implies that the engine is most productive in the
classical limit $q\to 0$.

We also study the $q$-dependence of the power 
$\mathcal{P}=\mathcal{W}/\tau$ where 
$\tau=t_{{\rm I}_1}+t_{{\rm I}_2}+t_{{\rm A}_1}+t_{{\rm A}_2}$ is the engine
cycle time.
The extracted work is independent of $t_{{\rm A}_1}$ and $t_{{\rm A}_2}$. 
Thus, for the optimal power, we  will set $t_{{\rm A}_1}=t_{{\rm A}_2}=0$ and
$t_{{\rm I}_1}=t_{{\rm I}_2}=\tau/2$. Then, the extracted work per cycle 
and the average power are given by
\begin{equation}
\begin{aligned}
\mathcal{W} & = \mathcal{W}_{\rm max} \frac{ \qty(1-e^{-\gamma\tau/2})^2}{
1-e^{-\gamma\tau}} \\
\mathcal{P} & = \mathcal{P}_{\rm max} \frac{ 4\qty(1-e^{-\gamma\tau/2})^2}{
\gamma\tau(1-e^{-\gamma\tau})} 
\end{aligned}
\end{equation}
with 
\begin{equation}
\mathcal{P}_{\rm max} = \frac{\gamma \mathcal{W}_{\rm max}}{4} .
\end{equation}
The power decreases monotonically as $\tau$ increases. 
It takes the maximum value $\mathcal{P}_{\rm max}$ in the $\tau\to 0$ limit.
Note that the maximum power is proportional to $\mathcal{W}_{\rm max}$.
Thus, the maximum power is a monotonically decreasing function of $q$. 

These results suggest that the quantum effect suppresses 
the power of the heat engine.
We note that a quantum coherence effect is absent in the quantum Otto engine
model considered in this work.
The Lindblad dynamics during the isochoric process and the simple form of
the time-dependent Hamiltonian satisfying
$\comm*{\hat{H}(t)}{\hat{H}(t')}=0$ during the adiabatic process do not
generate a quantum coherence~\cite{breuer2002theory}.
Thus, the quantum effect comes into play only through the discreteness 
of the energy level of the engine system.
In the classical limit with $q\ll 1$, the energy gap is smaller than the
thermal energy so that the heat flows freely between the system and the
reservoir.
However, in the quantum regime with $q\gg 1$, the discreteness of the energy
gap obstructs the heat flow, which makes the heat engine less efficient.
Recently, there was a report that the quantum coherence can enhance the
power of the heat engine~\cite{Klatzow:2019Experimental}. It would be
interesting to investigate the effects of the discreteness of the energy gap
and the quantum coherence simultaneously, which is beyond the scope of the
current work.

\section{Summary and discussions}\label{conclusion}

We have investigated the thermodynamic properties of the 
quantum Otto heat engine consisting of a harmonic oscillator.
The quantum system can be mapped to a classical thermodynamic 
system with the help of the quasi-probability distribution. 
Applying the second law of thermodynamics to the effective classical system, 
we have obtained the quantum mechanical bound for the
efficiency. The Q-function leads to the inequality that $\eta \leq \eta_Q$ with
the $\hbar$-dependent quantum mechanical bound $\eta_Q$. The equality holds
in the low temperature limit where $k_B T_i \ll \hbar\omega_i$.
Surprisingly, $\eta_Q \leq \eta_C$ so that 
the quantum mechanical bound provides a tighter bound
than the Carnot efficiency.

We also investigated the work and power of our engine model.
The work per engine cycle takes the maximum value in the limit
where the time intervals of the isothermal processes
tend to infinity.
The maximum value decreases as $q$ increases.
Thus, the engine produces the maximal work in the classical limit $q\to 0$.
In contrast to the work, the power takes the maximum in the small cycle 
time limit $\tau \rightarrow 0$.

\begin{figure}
\includegraphics*[width=\columnwidth]{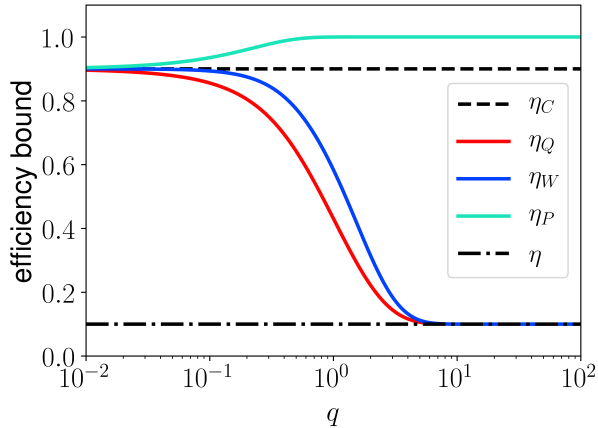}
\caption{
The bounds for the efficiency of the quantum Otto engine are plotted as
a function of $q$ with fixed $r_\omega = 0.9$ and $r_T = 0.1$. 
The engine efficiency is $\eta = 0.1$ 
(indicated by the black dashed-dotted line)
and the corresponding Carnot efficiency is $\eta_C = 0.9$
(indicated by the black dashed line).
The red, blue, and cyan solid lines represent the bounds $\eta_Q$, $\eta_W$,
and $\eta_P$, respectively.
} 
\label{fig_eta}
\end{figure}

One can consider the other quasi-probability distributions such as the
P-representation and the Wigner function instead of the Q-function. 
These choices only modify the effective temperature $\tilde{T}$.
That is, $\tilde{T} = \hbar\omega \bar{n}$
for the P-representation and $\hbar\omega\qty(\bar{n}+1/2)$ for the Wigner
function, while $\tilde{T} = \hbar\omega(\bar{n}+1)$ for the Q-function
as shown in \eqref{effective_parameter}. They yield the additional bounds
\begin{equation}
\begin{aligned}
\eta_P &= 1 - \frac{\omega_2 \bar{n}_2}{\omega_1 \bar{n}_1} \\
\eta_W &= 1 - \frac{\omega_2 \qty(\bar{n}_2+1/2)}{\omega_1
\qty(\bar{n}_1+1/2)}  .
\end{aligned}
\end{equation}
All the bounds satisfy the inequality
\begin{equation}\label{order_bounds}
    \eta \leq \eta_Q \leq \eta_W \leq \eta_C \leq \eta_P.
\end{equation}
They are compared in Fig.~\ref{fig_eta}.
Note that $\eta_P$ is larger than the Carnot efficiency and does not
provide useful information.
On the other hand, $\eta_W$ is smaller than the Carnot efficiency, but 
larger than $\eta_Q$.
The Q-function provides the most useful bound for the efficiency.
It may be interesting to find another quasi-probability distribution
leading to a tighter bound.

The exact mapping to the classical thermodynamic systems described by the
Fokker-Planck equation is possible only for the harmonic oscillator system.
Nevertheless, we expect that the similar quantum mechanical bound may exist
for other quantum heat engines.
For example, our system reduces to a two-level system
in the low temperature limit.
Since our formalism is still valid in that limit,
we expect that the efficiency of the quantum Otto heat engine with
the two-level system
would be bounded by the quantum mechanical bound. We leave the extension to
other quantum heat engines for future studies.

\begin{acknowledgements}
This work was supported by the National Research Foundation of Korea (NRF) grant funded by the Korea government (MSIP) (No. 2016R1A2B2013972).
S. L. acknowledges the support of National Research Foundation of Korea (NRF-2017R1A2B3006930).
\end{acknowledgements}

\appendix
\bibliographystyle{apsrev}
\bibliography{paper}

\begin{thebibliography}{35}
\expandafter\ifx\csname natexlab\endcsname\relax\def\natexlab#1{#1}\fi
\expandafter\ifx\csname bibnamefont\endcsname\relax
  \def\bibnamefont#1{#1}\fi
\expandafter\ifx\csname bibfnamefont\endcsname\relax
  \def\bibfnamefont#1{#1}\fi
\expandafter\ifx\csname citenamefont\endcsname\relax
  \def\citenamefont#1{#1}\fi
\expandafter\ifx\csname url\endcsname\relax
  \def\url#1{\texttt{#1}}\fi
\expandafter\ifx\csname urlprefix\endcsname\relax\def\urlprefix{URL }\fi
\providecommand{\bibinfo}[2]{#2}
\providecommand{\eprint}[2][]{\url{#2}}

\bibitem[{\citenamefont{Kittel and Kroemer}(1980)}]{Kittel:1980tpb}
\bibinfo{author}{\bibfnamefont{C.}~\bibnamefont{Kittel}} \bibnamefont{and}
  \bibinfo{author}{\bibfnamefont{H.}~\bibnamefont{Kroemer}},
  \emph{\bibinfo{title}{Thermal physics}} (\bibinfo{publisher}{W. H. Freeman},
  \bibinfo{address}{New York}, \bibinfo{year}{1980}), \bibinfo{edition}{2nd}
  ed.

\bibitem[{\citenamefont{Alicki}(1979)}]{alicki1979quantum}
\bibinfo{author}{\bibfnamefont{R.}~\bibnamefont{Alicki}}, \bibinfo{journal}{J.
  Phys. A: Math. Gen.} \textbf{\bibinfo{volume}{12}}, \bibinfo{pages}{L103}
  (\bibinfo{year}{1979}).

\bibitem[{\citenamefont{Kosloff}(1984)}]{kosloff1984quantum}
\bibinfo{author}{\bibfnamefont{R.}~\bibnamefont{Kosloff}}, \bibinfo{journal}{J.
  Chem. Phys.} \textbf{\bibinfo{volume}{80}}, \bibinfo{pages}{1625}
  (\bibinfo{year}{1984}).

\bibitem[{\citenamefont{Uzdin et~al.}(2015)\citenamefont{Uzdin, Levy, and
  Kosloff}}]{uzdin2015equivalence}
\bibinfo{author}{\bibfnamefont{R.}~\bibnamefont{Uzdin}},
  \bibinfo{author}{\bibfnamefont{A.}~\bibnamefont{Levy}}, \bibnamefont{and}
  \bibinfo{author}{\bibfnamefont{R.}~\bibnamefont{Kosloff}},
  \bibinfo{journal}{Phys. Rev. X} \textbf{\bibinfo{volume}{5}},
  \bibinfo{pages}{031044} (\bibinfo{year}{2015}).

\bibitem[{\citenamefont{Kosloff and Rezek}(2017)}]{kosloff2017quantum}
\bibinfo{author}{\bibfnamefont{R.}~\bibnamefont{Kosloff}} \bibnamefont{and}
  \bibinfo{author}{\bibfnamefont{Y.}~\bibnamefont{Rezek}},
  \bibinfo{journal}{Entropy} \textbf{\bibinfo{volume}{19}},
  \bibinfo{pages}{136} (\bibinfo{year}{2017}).

\bibitem[{\citenamefont{Lin and Chen}(2003)}]{lin2003performance}
\bibinfo{author}{\bibfnamefont{B.}~\bibnamefont{Lin}} \bibnamefont{and}
  \bibinfo{author}{\bibfnamefont{J.}~\bibnamefont{Chen}},
  \bibinfo{journal}{Phys. Rev. E} \textbf{\bibinfo{volume}{67}},
  \bibinfo{pages}{046105} (\bibinfo{year}{2003}).

\bibitem[{\citenamefont{Bender et~al.}(2000)\citenamefont{Bender, Brody, and
  Meister}}]{bender2000quantum}
\bibinfo{author}{\bibfnamefont{C.~M.} \bibnamefont{Bender}},
  \bibinfo{author}{\bibfnamefont{D.~C.} \bibnamefont{Brody}}, \bibnamefont{and}
  \bibinfo{author}{\bibfnamefont{B.~K.} \bibnamefont{Meister}},
  \bibinfo{journal}{J. Phys. A: Math. Gen.} \textbf{\bibinfo{volume}{33}},
  \bibinfo{pages}{4427} (\bibinfo{year}{2000}).

\bibitem[{\citenamefont{Kieu}(2006)}]{kieu2006quantum}
\bibinfo{author}{\bibfnamefont{T.~D.} \bibnamefont{Kieu}},
  \bibinfo{journal}{Eur. Phys. J. D} \textbf{\bibinfo{volume}{39}},
  \bibinfo{pages}{115} (\bibinfo{year}{2006}).

\bibitem[{\citenamefont{Kieu}(2004)}]{kieu2004second}
\bibinfo{author}{\bibfnamefont{T.~D.} \bibnamefont{Kieu}},
  \bibinfo{journal}{Phys. Rev. Lett.} \textbf{\bibinfo{volume}{93}},
  \bibinfo{pages}{140403} (\bibinfo{year}{2004}).

\bibitem[{\citenamefont{Abah et~al.}(2012)\citenamefont{Abah, Ro\ss{}nagel,
  Jacob, Deffner, Schmidt-Kaler, Singer, and Lutz}}]{Abah:2012Single-ion}
\bibinfo{author}{\bibfnamefont{O.}~\bibnamefont{Abah}},
  \bibinfo{author}{\bibfnamefont{J.}~\bibnamefont{Ro\ss{}nagel}},
  \bibinfo{author}{\bibfnamefont{G.}~\bibnamefont{Jacob}},
  \bibinfo{author}{\bibfnamefont{S.}~\bibnamefont{Deffner}},
  \bibinfo{author}{\bibfnamefont{F.}~\bibnamefont{Schmidt-Kaler}},
  \bibinfo{author}{\bibfnamefont{K.}~\bibnamefont{Singer}}, \bibnamefont{and}
  \bibinfo{author}{\bibfnamefont{E.}~\bibnamefont{Lutz}},
  \bibinfo{journal}{Phys. Rev. Lett.} \textbf{\bibinfo{volume}{109}},
  \bibinfo{pages}{203006} (\bibinfo{year}{2012}).

\bibitem[{\citenamefont{Ro{\ss}nagel et~al.}(2014)\citenamefont{Ro{\ss}nagel,
  Abah, Schmidt-Kaler, Singer, and Lutz}}]{rossnagel2014nanoscale}
\bibinfo{author}{\bibfnamefont{J.}~\bibnamefont{Ro{\ss}nagel}},
  \bibinfo{author}{\bibfnamefont{O.}~\bibnamefont{Abah}},
  \bibinfo{author}{\bibfnamefont{F.}~\bibnamefont{Schmidt-Kaler}},
  \bibinfo{author}{\bibfnamefont{K.}~\bibnamefont{Singer}}, \bibnamefont{and}
  \bibinfo{author}{\bibfnamefont{E.}~\bibnamefont{Lutz}},
  \bibinfo{journal}{Phys. Rev. Lett.} \textbf{\bibinfo{volume}{112}},
  \bibinfo{pages}{030602} (\bibinfo{year}{2014}).

\bibitem[{\citenamefont{Scully et~al.}(2003)\citenamefont{Scully, Zubairy,
  Agarwal, and Walther}}]{scully2003extracting}
\bibinfo{author}{\bibfnamefont{M.~O.} \bibnamefont{Scully}},
  \bibinfo{author}{\bibfnamefont{M.~S.} \bibnamefont{Zubairy}},
  \bibinfo{author}{\bibfnamefont{G.~S.} \bibnamefont{Agarwal}},
  \bibnamefont{and} \bibinfo{author}{\bibfnamefont{H.}~\bibnamefont{Walther}},
  \bibinfo{journal}{Science} \textbf{\bibinfo{volume}{299}},
  \bibinfo{pages}{862} (\bibinfo{year}{2003}).

\bibitem[{\citenamefont{Abah and Lutz}(2014)}]{abah2014efficiency}
\bibinfo{author}{\bibfnamefont{O.}~\bibnamefont{Abah}} \bibnamefont{and}
  \bibinfo{author}{\bibfnamefont{E.}~\bibnamefont{Lutz}},
  \bibinfo{journal}{Europhys. Lett.} \textbf{\bibinfo{volume}{106}},
  \bibinfo{pages}{20001} (\bibinfo{year}{2014}).

\bibitem[{\citenamefont{Niedenzu et~al.}(2016)\citenamefont{Niedenzu,
  Gelbwaser-Klimovsky, Kofman, and Kurizki}}]{Wolfgang:2016operation}
\bibinfo{author}{\bibfnamefont{W.}~\bibnamefont{Niedenzu}},
  \bibinfo{author}{\bibfnamefont{D.}~\bibnamefont{Gelbwaser-Klimovsky}},
  \bibinfo{author}{\bibfnamefont{A.~G.} \bibnamefont{Kofman}},
  \bibnamefont{and} \bibinfo{author}{\bibfnamefont{G.}~\bibnamefont{Kurizki}},
  \bibinfo{journal}{New J. Phys.} \textbf{\bibinfo{volume}{18}},
  \bibinfo{pages}{083012} (\bibinfo{year}{2016}).

\bibitem[{\citenamefont{Diaz de~la Cruz and
  Martin-Delgado}(2014)}]{Diaz:2014Quantum-information}
\bibinfo{author}{\bibfnamefont{J.~M.} \bibnamefont{Diaz de~la Cruz}}
  \bibnamefont{and} \bibinfo{author}{\bibfnamefont{M.~A.}
  \bibnamefont{Martin-Delgado}}, \bibinfo{journal}{Phys. Rev. A}
  \textbf{\bibinfo{volume}{89}}, \bibinfo{pages}{032327}
  (\bibinfo{year}{2014}).

\bibitem[{\citenamefont{Kosloff and Feldmann}(2002)}]{Kosloff:2002Discrete}
\bibinfo{author}{\bibfnamefont{R.}~\bibnamefont{Kosloff}} \bibnamefont{and}
  \bibinfo{author}{\bibfnamefont{T.}~\bibnamefont{Feldmann}},
  \bibinfo{journal}{Phys. Rev. E} \textbf{\bibinfo{volume}{65}},
  \bibinfo{pages}{055102} (\bibinfo{year}{2002}).

\bibitem[{\citenamefont{Deng et~al.}(2013)\citenamefont{Deng, Wang, Liu,
  H\"anggi, and Gong}}]{Jiawen:2013Boosting}
\bibinfo{author}{\bibfnamefont{J.}~\bibnamefont{Deng}},
  \bibinfo{author}{\bibfnamefont{Q.-h.} \bibnamefont{Wang}},
  \bibinfo{author}{\bibfnamefont{Z.}~\bibnamefont{Liu}},
  \bibinfo{author}{\bibfnamefont{P.}~\bibnamefont{H\"anggi}}, \bibnamefont{and}
  \bibinfo{author}{\bibfnamefont{J.}~\bibnamefont{Gong}},
  \bibinfo{journal}{Phys. Rev. E} \textbf{\bibinfo{volume}{88}},
  \bibinfo{pages}{062122} (\bibinfo{year}{2013}).

\bibitem[{\citenamefont{Campo et~al.}(2014)\citenamefont{Campo, Goold, and
  Paternostro}}]{Campo:2014more}
\bibinfo{author}{\bibfnamefont{A.~d.} \bibnamefont{Campo}},
  \bibinfo{author}{\bibfnamefont{J.}~\bibnamefont{Goold}}, \bibnamefont{and}
  \bibinfo{author}{\bibfnamefont{M.}~\bibnamefont{Paternostro}},
  \bibinfo{journal}{Sci. Rep.} \textbf{\bibinfo{volume}{4}},
  \bibinfo{pages}{6208} (\bibinfo{year}{2014}).

\bibitem[{\citenamefont{Santos et~al.}(2017)\citenamefont{Santos, Landi, and
  Paternostro}}]{santos2017wigner}
\bibinfo{author}{\bibfnamefont{J.~P.} \bibnamefont{Santos}},
  \bibinfo{author}{\bibfnamefont{G.~T.} \bibnamefont{Landi}}, \bibnamefont{and}
  \bibinfo{author}{\bibfnamefont{M.}~\bibnamefont{Paternostro}},
  \bibinfo{journal}{Phys. Rev. Lett.} \textbf{\bibinfo{volume}{118}},
  \bibinfo{pages}{220601} (\bibinfo{year}{2017}).

\bibitem[{\citenamefont{Gardiner and Zoller}(2004)}]{gardiner2004quantum}
\bibinfo{author}{\bibfnamefont{C.~W.} \bibnamefont{Gardiner}} \bibnamefont{and}
  \bibinfo{author}{\bibfnamefont{P.}~\bibnamefont{Zoller}},
  \emph{\bibinfo{title}{Quantum noise: a handbook of Markovian and
  non-Markovian quantum stochastic methods with applications to quantum
  optics}} (\bibinfo{publisher}{Springer Science}, \bibinfo{address}{Berlin,
  Heidelberg, New York}, \bibinfo{year}{2004}), \bibinfo{edition}{3rd} ed.

\bibitem[{\citenamefont{Seifert}(2005)}]{seifert:2005Entropy}
\bibinfo{author}{\bibfnamefont{U.}~\bibnamefont{Seifert}},
  \bibinfo{journal}{Phys. Rev. Lett.} \textbf{\bibinfo{volume}{95}},
  \bibinfo{pages}{040602} (\bibinfo{year}{2005}).

\bibitem[{\citenamefont{Wang et~al.}(2007)\citenamefont{Wang, He, and
  Mao}}]{Wang:2007Performance}
\bibinfo{author}{\bibfnamefont{J.}~\bibnamefont{Wang}},
  \bibinfo{author}{\bibfnamefont{J.}~\bibnamefont{He}}, \bibnamefont{and}
  \bibinfo{author}{\bibfnamefont{Z.}~\bibnamefont{Mao}}, \bibinfo{journal}{Sci.
  China Ser. G: Phys. Mech. Astron.} \textbf{\bibinfo{volume}{50}},
  \bibinfo{pages}{163} (\bibinfo{year}{2007}).

\bibitem[{\citenamefont{Liu et~al.}(2009)\citenamefont{Liu, Chen, Wu, and
  Sun}}]{Liu2009:Ecological}
\bibinfo{author}{\bibfnamefont{X.}~\bibnamefont{Liu}},
  \bibinfo{author}{\bibfnamefont{L.}~\bibnamefont{Chen}},
  \bibinfo{author}{\bibfnamefont{F.}~\bibnamefont{Wu}}, \bibnamefont{and}
  \bibinfo{author}{\bibfnamefont{F.}~\bibnamefont{Sun}}, \bibinfo{journal}{Sci.
  China Ser. G: Phys. Mech. Astron.} \textbf{\bibinfo{volume}{52}},
  \bibinfo{pages}{1976} (\bibinfo{year}{2009}).

\bibitem[{\citenamefont{Breuer et~al.}(2002)\citenamefont{Breuer, Petruccione
  et~al.}}]{breuer2002theory}
\bibinfo{author}{\bibfnamefont{H.-P.} \bibnamefont{Breuer}},
  \bibinfo{author}{\bibfnamefont{F.}~\bibnamefont{Petruccione}},
  \bibnamefont{et~al.}, \emph{\bibinfo{title}{The theory of open quantum
  systems}} (\bibinfo{publisher}{Oxford University Press},
  \bibinfo{address}{Berlin}, \bibinfo{year}{2002}).

\bibitem[{\citenamefont{Spohn and Lebowitz}(1978)}]{Spohn:1978Irreversible}
\bibinfo{author}{\bibfnamefont{H.}~\bibnamefont{Spohn}} \bibnamefont{and}
  \bibinfo{author}{\bibfnamefont{J.~L.} \bibnamefont{Lebowitz}},
  \bibinfo{journal}{Adv. Chem. Phys.} \textbf{\bibinfo{volume}{38}},
  \bibinfo{pages}{109} (\bibinfo{year}{1978}).

\bibitem[{\citenamefont{Santos et~al.}(2018{\natexlab{a}})\citenamefont{Santos,
  C{\'e}leri, Brito, Landi, and Paternostro}}]{santos2018spin}
\bibinfo{author}{\bibfnamefont{J.~P.} \bibnamefont{Santos}},
  \bibinfo{author}{\bibfnamefont{L.~C.} \bibnamefont{C{\'e}leri}},
  \bibinfo{author}{\bibfnamefont{F.}~\bibnamefont{Brito}},
  \bibinfo{author}{\bibfnamefont{G.~T.} \bibnamefont{Landi}}, \bibnamefont{and}
  \bibinfo{author}{\bibfnamefont{M.}~\bibnamefont{Paternostro}},
  \bibinfo{journal}{Phys. Rev. A} \textbf{\bibinfo{volume}{97}},
  \bibinfo{pages}{052123} (\bibinfo{year}{2018}{\natexlab{a}}).

\bibitem[{\citenamefont{Santos et~al.}(2018{\natexlab{b}})\citenamefont{Santos,
  de~Paula~Jr, Drumond, Landi, and Paternostro}}]{santos2018irreversibility}
\bibinfo{author}{\bibfnamefont{J.~P.} \bibnamefont{Santos}},
  \bibinfo{author}{\bibfnamefont{A.~L.} \bibnamefont{de~Paula~Jr}},
  \bibinfo{author}{\bibfnamefont{R.}~\bibnamefont{Drumond}},
  \bibinfo{author}{\bibfnamefont{G.~T.} \bibnamefont{Landi}}, \bibnamefont{and}
  \bibinfo{author}{\bibfnamefont{M.}~\bibnamefont{Paternostro}},
  \bibinfo{journal}{Phys. Rev. A} \textbf{\bibinfo{volume}{97}},
  \bibinfo{pages}{050101} (\bibinfo{year}{2018}{\natexlab{b}}).

\bibitem[{\citenamefont{Husimi}(1940)}]{husimi1940some}
\bibinfo{author}{\bibfnamefont{K.}~\bibnamefont{Husimi}},
  \bibinfo{journal}{Proc. Phys. Math. Soc. Jpn.} \textbf{\bibinfo{volume}{22}},
  \bibinfo{pages}{264} (\bibinfo{year}{1940}).

\bibitem[{\citenamefont{Blasiak et~al.}(2007)\citenamefont{Blasiak, Horzela,
  Penson, Solomon, and Duchamp}}]{Blasiak:2007Combinatorics}
\bibinfo{author}{\bibfnamefont{P.}~\bibnamefont{Blasiak}},
  \bibinfo{author}{\bibfnamefont{A.}~\bibnamefont{Horzela}},
  \bibinfo{author}{\bibfnamefont{K.~A.} \bibnamefont{Penson}},
  \bibinfo{author}{\bibfnamefont{A.~I.} \bibnamefont{Solomon}},
  \bibnamefont{and} \bibinfo{author}{\bibfnamefont{G.~H.~E.}
  \bibnamefont{Duchamp}}, \bibinfo{journal}{Am. J. Phys.}
  \textbf{\bibinfo{volume}{75}}, \bibinfo{pages}{639} (\bibinfo{year}{2007}).

\bibitem[{\citenamefont{Gardiner}(2010)}]{Gardiner:2010stochastic}
\bibinfo{author}{\bibfnamefont{C.}~\bibnamefont{Gardiner}},
  \emph{\bibinfo{title}{{Stochastic Methods: A Handbook for the Natural and
  Social Sciences}}} (\bibinfo{publisher}{Springer}, \bibinfo{address}{New
  York}, \bibinfo{year}{2010}), \bibinfo{edition}{4th} ed.

\bibitem[{\citenamefont{Risken}(1996)}]{Risken:1996FokkerPlanck}
\bibinfo{author}{\bibfnamefont{H.}~\bibnamefont{Risken}},
  \emph{\bibinfo{title}{{The Fokker-Planck Equation: Methods of Solution and
  Applications}}} (\bibinfo{publisher}{Springer-Verlag},
  \bibinfo{address}{Berlin}, \bibinfo{year}{1996}), \bibinfo{edition}{2nd} ed.

\bibitem[{\citenamefont{Van~Kampen}(2011)}]{VanKampen:2011Stochastic}
\bibinfo{author}{\bibfnamefont{N.~G.} \bibnamefont{Van~Kampen}},
  \emph{\bibinfo{title}{{Stochastic Processes in Physics and Chemistry}}}
  (\bibinfo{publisher}{Elsevier}, \bibinfo{address}{New York},
  \bibinfo{year}{2011}), \bibinfo{edition}{3rd} ed.

\bibitem[{\citenamefont{Wehrl}(1979)}]{wehrl1979relation}
\bibinfo{author}{\bibfnamefont{A.}~\bibnamefont{Wehrl}}, \bibinfo{journal}{Rep.
  Math. Phys.} \textbf{\bibinfo{volume}{16}}, \bibinfo{pages}{353}
  (\bibinfo{year}{1979}).

\bibitem[{\citenamefont{Wehrl}(1978)}]{wehrl1978general}
\bibinfo{author}{\bibfnamefont{A.}~\bibnamefont{Wehrl}}, \bibinfo{journal}{Rev.
  Mod. Phys.} \textbf{\bibinfo{volume}{50}}, \bibinfo{pages}{221}
  (\bibinfo{year}{1978}).

\bibitem[{\citenamefont{Klatzow et~al.}(2019)\citenamefont{Klatzow, Becker,
  Ledingham, Weinzetl, Kaczmarek, Saunders, Nunn, Walmsley, Uzdin, and
  Poem}}]{Klatzow:2019Experimental}
\bibinfo{author}{\bibfnamefont{J.}~\bibnamefont{Klatzow}},
  \bibinfo{author}{\bibfnamefont{J.~N.} \bibnamefont{Becker}},
  \bibinfo{author}{\bibfnamefont{P.~M.} \bibnamefont{Ledingham}},
  \bibinfo{author}{\bibfnamefont{C.}~\bibnamefont{Weinzetl}},
  \bibinfo{author}{\bibfnamefont{K.~T.} \bibnamefont{Kaczmarek}},
  \bibinfo{author}{\bibfnamefont{D.~J.} \bibnamefont{Saunders}},
  \bibinfo{author}{\bibfnamefont{J.}~\bibnamefont{Nunn}},
  \bibinfo{author}{\bibfnamefont{I.~A.} \bibnamefont{Walmsley}},
  \bibinfo{author}{\bibfnamefont{R.}~\bibnamefont{Uzdin}}, \bibnamefont{and}
  \bibinfo{author}{\bibfnamefont{E.}~\bibnamefont{Poem}},
  \bibinfo{journal}{Phys. Rev. Lett.} \textbf{\bibinfo{volume}{122}},
  \bibinfo{pages}{110601} (\bibinfo{year}{2019}).

\end{thebibliography}

\end{document}